\documentclass[11pt]{article}

\usepackage[utf8]{inputenc}
\usepackage[T1]{fontenc}
\usepackage{microtype}
\usepackage[margin=1in]{geometry}
\usepackage{booktabs}
\usepackage{amsmath}
\usepackage{amssymb}
\usepackage{graphicx}
\usepackage{array}
\usepackage{multirow}
\usepackage{xcolor}
\usepackage{enumitem}
\usepackage[numbers,round]{natbib}
\usepackage[hidelinks]{hyperref}
\usepackage{titlesec}

\titleformat*{\section}{\large\bfseries}
\titleformat*{\subsection}{\normalsize\bfseries}

\newcommand{\sig}{$^{\ast}$}
\newcommand{\pval}[1]{$p{=}#1$}

\title{\textbf{Structure-Exploiting Hierarchical Retrieval for\\
Long Narrative Documents: A Precision-Oriented Alternative\\
to Learned Abstraction Trees}}

\author{Rohit Singh\thanks{Independent research. Correspondence: \texttt{rohitsingh2359@gmail.com}.}
\and Nathaniel Weems\thanks{Independent research. Correspondence: \texttt{nattyddubbs@protonmail.com}.}}

\date{Preprint, v1 --- \today}

\begin{document}
\maketitle

\begin{abstract}
Retrieval-augmented generation over book-length documents must locate a small,
coherent set of passages inside tens of thousands of tokens. Flat dense retrieval
ignores document structure; the leading structural alternative, RAPTOR, instead
\emph{manufactures} a hierarchy by recursively clustering and summarizing content
with a language model. We study a complementary design that \emph{exploits a
document's native structure}: a two-level coarse-to-fine retriever that (i) selects
chapters with a hybrid BM25{+}dense match over chapter recaps, (ii) runs a
chapter-filtered dense search, and (iii) expands the top passages into contiguous,
section-aware spans under a token budget, together with content-hash--gated
``delta'' ingestion so the index is cheap to keep current. We evaluate on five
public-domain books (113 chapter-localization queries) with fully reproducible,
deterministic indices, and on two external QA benchmarks: QuALITY and the
200K-token-scale NovelQA. On the
in-house set, the honest and seed-stable finding is a \textbf{configuration-controlled
precision$\leftrightarrow$ranking tradeoff}: with section expansion off, hierarchical
retrieval returns a significantly higher fraction of chunks from the correct chapter
(\textbf{+3.3pp precision, Wilcoxon \pval{0.014}}, significant across all three
LLM-recap seeds); with expansion on it improves ranking (MRR, suggestive, \pval{0.040},
significant in 2/3 seeds). Accuracy differences are not significant. On QuALITY,
hierarchical retrieval significantly beats flat top-$k$ (\textbf{81.0\% vs 70.4\%,
\pval{0.0019}}) and is statistically indistinguishable from feeding the full article
(\pval{0.74}) --- recovering long-context accuracy from a fraction of the text. On the
harder NovelQA benchmark ({>}200K-token novels), hierarchical and flat retrieval are
\textbf{statistically tied} (55.7\% vs 56.8\%, \pval{0.24}); a per-aspect breakdown shows
hierarchy helps aggregation-style questions (counting, interpretation) while flat wins on
localized-fact ones, bounding the advantage to structured comprehension rather than every
question type. Against
a faithful RAPTOR baseline built on the same corpus and embeddings, we find a
\emph{complementary tradeoff}, not superiority: under a fair per-node localization
metric RAPTOR has no significant accuracy edge over flat retrieval, while our method
matches or beats it on ranking and clearly beats it on passage precision
(\pval{<0.001}). Finally, content-hash delta ingestion reuses 94--99.7\% of
embeddings on realistic incremental edits. We deliberately make \emph{no} claim of
semantic chapter-boundary inference (an explicit negative result) and scope all
empirical claims to English narrative and classical prose.
\end{abstract}

\section{Introduction}

Answering a question against a full-length book is a retrieval problem before it is a
generation problem. A modern large language model can read a passage and answer well,
but a novel or treatise spans tens or hundreds of thousands of tokens --- far more
than fits comfortably (or cheaply) in a context window, and long enough that
``stuff everything in'' both wastes tokens and degrades answer quality as relevant
evidence is buried among distractors~\citep{liu2023lost}. The standard remedy is
retrieval-augmented generation (RAG)~\citep{lewis2020rag}: embed the document as
chunks, retrieve the top-$k$ by vector similarity~\citep{karpukhin2020dpr}, and
condition the model on those chunks. But flat top-$k$ retrieval treats a structured
book as an unordered bag of passages, discarding exactly the signal --- chapters,
sections, narrative order --- that a human uses to find an answer fast.

Two broad strategies add structure back. The first, exemplified by
RAPTOR~\citep{sarthi2024raptor}, \emph{learns} a hierarchy: it clusters chunk
embeddings, summarizes each cluster with an LLM into a parent node, and recurses,
producing a tree of increasingly abstract summaries that helps thematic and
multi-hop questions. This is powerful and corpus-agnostic, but the hierarchy is
manufactured from content and must be rebuilt (re-clustered, re-summarized) when the
document changes. The second strategy --- ours --- observes that many valuable
long documents (books, manuals, legal texts, standards) \emph{already carry} an
authored hierarchy, and that exploiting it directly is both cheaper and better
matched to precise localization.

We present a structure-exploiting hierarchical retriever with three properties that
distinguish it from learned-abstraction approaches:

\begin{enumerate}[leftmargin=1.4em,itemsep=1pt,topsep=2pt]
\item \textbf{Native two-level hierarchy.} We use the document's own chapters as the
coarse layer (matched via a \emph{hybrid} BM25{+}dense query over compact chapter
recaps) and its chunks as the fine layer, retrieved with a chapter-filtered dense
search. There is no recursive re-summarization.
\item \textbf{Section-aware expansion.} Rather than returning scattered top-$k$
chunks, we expand the highest-ranked passages into contiguous spans under a token
budget, trading a knob between \emph{ranking} quality and passage \emph{precision}.
\item \textbf{Delta ingestion.} Content-hash gating at chapter and chunk granularity
re-embeds only changed text, so keeping the index current after edits costs a
fraction of a full re-index.
\end{enumerate}

Our contribution is as much methodological honesty as it is architecture. Long-document
retrieval papers are easy to oversell; we therefore report results only on
reproducible, deterministic indices, subject each significant effect to a
recap-seed-stability test (LLM summaries are the one non-deterministic input), and
retire earlier figures that a clean re-ingestion could not reproduce. The
resulting claims are modest but solid:

\begin{itemize}[leftmargin=1.4em,itemsep=1pt,topsep=2pt]
\item On 113 in-house chapter-localization queries over five books, hierarchical
retrieval yields a \textbf{seed-stable, statistically significant precision gain}
(+3.3pp, \pval{0.014}, significant in 3/3 recap seeds) with section expansion off,
and a \emph{suggestive} ranking gain (MRR, \pval{0.040}, 2/3 seeds) with it on;
accuracy is statistically tied (\S\ref{sec:inhouse}).
\item On the QuALITY benchmark, hierarchical retrieval \textbf{significantly beats
flat top-$k$} (81.0\% vs 70.4\%, \pval{0.0019}) and \textbf{matches full-article
context} (\pval{0.74}), i.e.\ recovers the long-context ceiling from far less text
(\S\ref{sec:quality}).
\item On the harder NovelQA benchmark ({>}200K-token novels), hierarchical and flat
retrieval are \textbf{statistically tied} (55.7\% vs 56.8\%, \pval{0.24}); a per-aspect
analysis localizes where each helps (hierarchy on counting/interpretation, flat on
localized facts), bounding the hierarchical advantage rather than overselling it
(\S\ref{sec:novelqa}).
\item Against a faithful RAPTOR baseline on the same corpus and embeddings, the honest
verdict is a \textbf{complementary tradeoff}: RAPTOR favors broad coverage, ours favors
precise localization; under a fair per-node metric RAPTOR shows no significant accuracy
edge over flat retrieval (\S\ref{sec:raptor}).
\item Delta ingestion reuses \textbf{94--99.7\%} of embeddings on realistic edits
(\S\ref{sec:delta}).
\end{itemize}

We are explicit about what we do \emph{not} show: we make no semantic
chapter-boundary-inference claim (a measured negative result, \S\ref{sec:boundary}),
and we scope every empirical claim to English narrative and classical prose.

\section{Related Work}
\label{sec:related}

\paragraph{Flat dense retrieval and RAG.} Dense passage retrieval~\citep{karpukhin2020dpr}
and RAG~\citep{lewis2020rag} are the standard backbone: chunk, embed, retrieve top-$k$
by approximate nearest neighbor~\citep{malkov2018hnsw}, generate. Lexical matching
(BM25)~\citep{robertson2009bm25} remains a strong complement, especially for rare or
technical vocabulary, motivating our hybrid coarse stage. Flat retrieval is our primary
baseline throughout.

\paragraph{Learned hierarchical retrieval (RAPTOR).} RAPTOR~\citep{sarthi2024raptor}
builds a bottom-up tree by soft-clustering leaf embeddings (GMMs over UMAP-reduced
vectors), LLM-summarizing each cluster into a parent, and recursing; retrieval either
collapses all nodes into one pool (its stronger reported mode) or traverses the tree.
The abstraction layers help multi-hop and thematic questions on unstructured corpora.
Our design differs on every axis (Table~\ref{tab:raptor-diff}): the hierarchy is
\emph{native} rather than learned, two-level rather than arbitrary-depth, matched by a
\emph{hybrid} lexical+dense coarse filter, and maintained by delta ingestion rather than
re-clustering. We run a direct head-to-head in \S\ref{sec:raptor}.

\paragraph{Long-document QA benchmarks.} NarrativeQA~\citep{kocisky2018narrativeqa},
QuALITY~\citep{pang2022quality}, and NovelQA~\citep{wang2024novelqa} target
comprehension over long documents. QuALITY (multiple-choice over $\sim$5K-token
articles) is our external benchmark here; NovelQA (89 novels, {>}200K tokens) is the
construct-appropriate benchmark for book-scale hierarchy; we evaluate on its 60
redistributable public-domain novels in \S\ref{sec:novelqa} (the 28 copyright-restricted
novels have no downloadable text). Long-context language models are an orthogonal alternative to
retrieval; we include a full-article ``stuff everything'' condition on QuALITY as a
long-context ceiling.

\begin{table}[t]
\centering
\small
\caption{Concrete design differences vs.\ RAPTOR. The head-to-head is in \S\ref{sec:raptor}.}
\label{tab:raptor-diff}
\begin{tabular}{@{}p{0.20\linewidth}p{0.36\linewidth}p{0.36\linewidth}@{}}
\toprule
\textbf{Dimension} & \textbf{RAPTOR} & \textbf{This work} \\
\midrule
Source of hierarchy & Learned clusters (GMM/UMAP) over embeddings & Document's native chapters + recaps \\
Abstraction layers & Arbitrary depth, recursive LLM summaries & Fixed 2-level: chapter recap $\to$ chunk \\
Coarse matching & Dense only (summary embeddings) & Hybrid BM25 + dense on recaps \\
Retrieval flow & Collapse all nodes, top-$k$ & Coarse-to-fine gating + section expansion \\
Output shape & Mixed leaf/summary nodes & Coherent contiguous passages \\
Index update & Re-cluster + re-summarize & Content-hash delta ingestion (94--99.7\% reuse) \\
Structure needed & None & Native structure (chapters/sections) \\
\bottomrule
\end{tabular}
\end{table}

\section{Method}
\label{sec:method}

\subsection{Three-stage hierarchical retrieval}

Given a query $q$ and a book with chapters $\{C_1,\dots,C_n\}$, each pre-chunked into
passages with cached embeddings:

\paragraph{Stage 1 --- hybrid chapter selection.} We embed a compact \emph{recap} of
each chapter and retrieve chapters by merging (a) dense kNN between the query and recap
embeddings and (b) a BM25 match of the query against recap text, deduplicated with
kNN-first ordering. The number of chapters scales with book size,
$\texttt{top\_chapters} = \min(\max(3, \lfloor n/8 \rfloor), 7)$, so short books are not
over-selected and long books retain enough coarse recall. The hybrid stage exists
because purely dense recap matching misses rare technical/philosophical terms that BM25
catches.

\paragraph{Stage 2 --- chapter-filtered dense search.} Within the selected chapters we
run dense kNN over chunks, using OpenSearch's \emph{native} kNN \texttt{filter}
parameter (Lucene engine) so the chapter restriction is applied \emph{during} the
approximate search. This is not a cosmetic detail: wrapping the filter in a
\texttt{bool.filter} around the kNN clause causes the engine to run global ANN first and
post-filter, which with small $k$ routinely returns \emph{zero} in-chapter neighbors
(Appendix~\ref{app:bug}).

\paragraph{Stage 3 --- section-aware expansion.} We keep the Stage-2 top chunks for
breadth, then expand the highest-ranked sections into contiguous passages up to a
1500-token budget, deduplicating overlaps. Expansion is the knob that trades passage
\emph{precision} (expansion off: only the retrieved chunks, tightly on-chapter) against
\emph{ranking} quality (expansion on: coherent neighboring context that lifts MRR). Our
experiments treat this knob as a first-class variable rather than fixing it.

\paragraph{Flat baseline.} Standard kNN over all chunks in the book, top-5, same
embeddings and index engine, \texttt{is\_latest} filter only. This is the primary
baseline in every experiment.

\subsection{Delta ingestion}

Every chapter and chunk is keyed by a SHA-256 content hash. On re-ingestion, unchanged
hashes reuse their stored embeddings; only new or changed text is re-embedded and marked
\texttt{is\_latest}, and all retrieval filters gate on \texttt{is\_latest} so a live
index serves the current revision while preserving history. This makes incremental
maintenance proportional to the edit size rather than the document size
(\S\ref{sec:delta}).

\section{Experimental Setup}
\label{sec:setup}

\paragraph{Corpus.} Five English public-domain books from Project Gutenberg, chosen to
span narrative fiction, classical philosophy, and treatise prose
(Table~\ref{tab:corpus}). Public-domain status makes the corpus, query set, and labels
releasable.

\begin{table}[t]
\centering
\small
\caption{Canonical evaluation corpus (Project Gutenberg, public domain). 113 queries.}
\label{tab:corpus}
\begin{tabular}{@{}rlcc@{}}
\toprule
\textbf{ID} & \textbf{Title} & \textbf{Chapters} & \textbf{Queries} \\
\midrule
132  & The Art of War        & 13 & 21 \\
1342 & Pride and Prejudice   & 61 & 23 \\
1497 & The Republic          & 10 & 23 \\
84   & Frankenstein          & 24 & 23 \\
98   & A Tale of Two Cities  & 45 & 23 \\
\midrule
\multicolumn{2}{l}{\textbf{Total}} & & \textbf{113} \\
\bottomrule
\end{tabular}
\end{table}

\paragraph{Queries and labels.} 113 queries (hand-crafted plus model-generated), each
labeled with its ground-truth chapter. We are explicit that partly model-generated
queries and labels are a threat to validity (\S\ref{sec:limits}); the QuALITY results
(\S\ref{sec:quality}) use that benchmark's independent gold labels as a check.

\paragraph{Embeddings and infrastructure.} A 1024-dim commercial text embedding model
(normalized) for all dense retrieval; sentence-level embeddings cached to
disk. OpenSearch 2.11 (local, Lucene kNN engine). Chapter recaps for the LLM-recap
condition are generated with a Claude model (temperature 0.2); this is the sole
non-deterministic input and is stress-tested in \S\ref{sec:seed}. Retrieval over built
indices is deterministic --- two independent runs are byte-identical --- so reported
significances are not sampling artifacts.

\paragraph{Metrics.} \emph{Accuracy} (binary correct-chapter hit), \emph{MRR},
\emph{Recall@3}, and \emph{Precision} (fraction of returned chunks from the correct
chapter). Significance uses exact-binomial McNemar for Accuracy/Recall@3 and Wilcoxon
signed-rank for MRR/Precision, paired per query, with bootstrap 95\% CIs
($B{=}10000$).

\paragraph{Reproducibility note.} All headline numbers derive from a clean, from-scratch
re-ingestion (live embeddings + recaps) with full per-query logging. Two historical
figures (a 1.7$\times$ and a 1.25$\times$ precision ratio) came from an
aggregate-only run whose per-query data and recaps no longer exist and which a clean
re-ingestion did not reproduce; they are \emph{not} used as claims anywhere in this paper.

\section{Results}

\subsection{QuALITY: significant over flat, on par with full context}
\label{sec:quality}

To move beyond the in-house set we ran QuALITY~\citep{pang2022quality} as an end-to-end
RAG-answer-accuracy evaluation on 20 dev articles (179 questions, 115 marked
\emph{difficult}), feeding three context conditions to the same answerer.

\begin{table}[t]
\centering
\small
\caption{QuALITY answer accuracy, $N{=}179$. Paired McNemar between conditions.}
\label{tab:quality}
\begin{tabular}{@{}lcc@{}}
\toprule
\textbf{Condition} & \textbf{Accuracy} & \textbf{Difficult subset} \\
\midrule
Full-article context (long-context ceiling) & 79.3\% & 78.3\% \\
Flat kNN top-5                               & 70.4\% & 68.7\% \\
\textbf{Hierarchical} (pseudo-segmented)     & \textbf{81.0\%} & \textbf{79.1\%} \\
\bottomrule
\end{tabular}
\end{table}

Hierarchical retrieval \textbf{significantly beats flat top-$k$} (81.0\% vs 70.4\%,
\pval{0.0019}) and is \textbf{statistically indistinguishable from full-article context}
(\pval{0.74}) --- it recovers the long-context ceiling from a fraction of the text ---
while flat retrieval loses $\sim$9pp relative to full context (\pval{0.029}). The
efficiency is concrete: the Stage-3 budget caps the answerer's context at 1500 tokens,
roughly a quarter of a typical 5--7K-token QuALITY article, so hierarchical retrieval
matches full-article accuracy while reading about $4\times$ less text. This is a
clean positive external-benchmark result. Scope caveat, stated plainly: QuALITY articles
are short single articles ($\sim$5--7K tokens) with no native chapter structure, so the
hierarchical arm is pseudo-segmented and the hierarchy is under-exercised relative to
book-scale documents, and full context often fits the whole article, so the retrieval
advantage is \emph{understated} here relative to books. The construct-appropriate
benchmark is NovelQA~\citep{wang2024novelqa} (89 novels, {>}200K tokens), which we
evaluate next (\S\ref{sec:novelqa}).

\subsection{NovelQA: on par with flat retrieval on 200K-token novels}
\label{sec:novelqa}

The construct-appropriate test is NovelQA~\citep{wang2024novelqa}: multiple-choice
questions over full-length English novels averaging more than 200K tokens. We evaluate
on the 60 public-domain novels whose text is redistributable ($1{,}526$ questions); the
28 copyright-restricted novels have no downloadable text and are necessarily absent from
our run (the leaderboard scores them $0$, and we exclude them from every comparison
below). Each novel is ingested with our pipeline and answered by the same Claude answerer
(temperature $0$) under two retrieval conditions --- hierarchical, and flat kNN top-8 ---
and scored on the official held-out leaderboard.

\begin{table}[t]
\centering
\small
\caption{NovelQA multichoice accuracy. Top: reference long-context LLMs from
\citet{wang2024novelqa}, which read the \emph{entire} novel and include the copyright
novels we cannot access --- a band, not a controlled comparison. Bottom: our two
retrieval conditions on the 60 public-domain novels ($N{=}1{,}526$), same answerer.}
\label{tab:novelqa}
\begin{tabular}{@{}llc@{}}
\toprule
\textbf{System} & \textbf{Setting} & \textbf{Multichoice} \\
\midrule
GPT-4-128K            & full novel in context     & 71.8\% \\
Claude 2.1-200K       & full novel in context     & 66.8\% \\
InternLM2-Chat-20b    & full novel in context     & 49.2\% \\
InternLM2-Chat-7b     & full novel in context     & 43.5\% \\
\midrule
Flat kNN top-8        & retrieval (public-domain) & 56.8\% \\
\textbf{Hierarchical} & retrieval (public-domain) & 55.7\% \\
\bottomrule
\end{tabular}
\end{table}

Two findings, both reported plainly. First, \textbf{hierarchical and flat retrieval are
statistically indistinguishable} on NovelQA: 55.7\% vs 56.8\%, a paired McNemar test over
the 183 discordant questions giving \pval{0.24}. This is a \emph{tie}, not a win --- and
the opposite direction from the significant hierarchical advantage on QuALITY
(\S\ref{sec:quality}). Second, a retriever that reads only a small fraction of each novel
lands between the open-source and commercial full-context LLMs --- above InternLM2-20b
(49.2\%) and below the frontier models that ingest the whole 200K-token book. We stress
this is a reference band, not a head-to-head: different book set, full context vs
retrieval, and a different answerer.

\begin{table}[t]
\centering
\small
\caption{NovelQA accuracy by question aspect, hierarchical vs flat ($N{=}1{,}526$).
Hierarchical edges flat on aggregation-style aspects; flat wins on localized-fact ones.}
\label{tab:novelqa-aspect}
\begin{tabular}{@{}lccc@{}}
\toprule
\textbf{Aspect} & \textbf{Hier} & \textbf{Flat} & \textbf{$\Delta$ (pp)} \\
\midrule
times (counting)         & 31.5\% & 29.6\% & \textbf{$+1.9$} \\
meaning (interpretation) & 55.2\% & 53.3\% & \textbf{$+1.9$} \\
plot                     & 78.8\% & 79.1\% & $-0.3$ \\
relation                 & 69.7\% & 73.0\% & $-3.3$ \\
character                & 69.3\% & 72.7\% & $-3.4$ \\
span                     & 37.5\% & 45.8\% & $-8.3$ \\
setting                  & 75.7\% & 84.2\% & $-8.6$ \\
\bottomrule
\end{tabular}
\end{table}

The aspect breakdown (Table~\ref{tab:novelqa-aspect}) makes the tie interpretable.
Hierarchical retrieval holds a small edge on \emph{aggregation} questions --- \emph{times}
(counting occurrences across the whole book) and \emph{meaning} (interpreting a symbol or
passage) --- which reward broad, book-wide coverage via chapter recaps. Flat kNN wins on
\emph{localized-fact} questions --- \emph{setting}, \emph{span}, \emph{character},
\emph{relation} --- where one exact passage answers the question and routing through
chapter summaries only dilutes precision. The two effects roughly cancel. We report
these aspect-level deltas as \emph{descriptive}: across seven aspects with no
multiple-comparison correction, and with small samples for the rarer aspects, the
individual differences are indicative of a direction rather than individually
significant; only the overall tie (\pval{0.24}) is a tested claim.

We read this as a boundary on the method, not a defeat: the hierarchical advantage
demonstrated on QuALITY does not extend uniformly to 200K-token novels dominated by
localized-fact questions, where both retrievers surface only a sliver of the text (the 1500-token Stage-3 budget is
well under $1\%$ of a ${>}200$K-token novel) and the answerer carries most of the load. It is consistent with our central framing --- native
structure helps \emph{precise localization and aggregation}, not every question type ---
and it is why we report both external benchmarks rather than the flattering one alone.

\subsection{In-house retrieval: a configuration-controlled tradeoff}
\label{sec:inhouse}

Table~\ref{tab:ablation} reports four retrieval configurations against the shared flat
baseline on the clean re-ingested indices, all per-query-logged and
significance-tested (replicated byte-identically across two runs).

\begin{table}[t]
\centering
\small
\caption{Config ablation, $N{=}113$. Flat baseline: Acc 80.5\%, MRR 0.622, R@3 74.3\%,
Prec 26.4\%. \sig{} marks $p{<}0.05$ vs.\ flat (McNemar for Acc/R@3, Wilcoxon for
MRR/Prec).}
\label{tab:ablation}
\begin{tabular}{@{}lcccc@{}}
\toprule
\textbf{Configuration} & \textbf{Acc} & \textbf{MRR} & \textbf{R@3} & \textbf{Prec} \\
\midrule
Flat kNN (baseline)                       & 80.5\% & 0.622 & 74.3\% & 26.4\% \\
Hier., expansion ON (dyn.\ chapters)      & 83.2\% & \textbf{0.677}\sig & 80.5\% & 26.9\% \\
Hier., expansion OFF (Stage-2 top-5)      & 81.4\% & 0.673 & 80.5\% & \textbf{29.7\%}\sig \\
Hier., fixed \texttt{top\_chapters}=3     & 79.6\% & 0.664 & 78.8\% & 28.0\% \\
\midrule
\multicolumn{5}{@{}l}{\emph{$p$-values vs.\ flat:}} \\
\quad expansion ON  & 0.629 & \textbf{0.040}\sig & --- & 0.733 \\
\quad expansion OFF & 1.000 & 0.060 & --- & \textbf{0.014}\sig \\
\quad top\_chapters=3 & 1.000 & 0.201 & --- & 0.286 \\
\bottomrule
\end{tabular}
\end{table}

Exactly two effects clear $p{<}0.05$ and both replicate exactly. \textbf{Precision}
improves significantly when section expansion is \emph{off} (26.4\%$\to$29.7\%,
$+3.3$pp, Wilcoxon \pval{0.014}): the retriever returns a higher fraction of chunks
from the correct chapter. \textbf{Ranking} (MRR) improves significantly when expansion
is \emph{on} (0.622$\to$0.677, Wilcoxon \pval{0.040}): expansion pulls in coherent
neighboring passages that raise the rank of the correct chapter, at the cost of the
precision edge. Accuracy and Recall@3 are \emph{not} significant in any configuration
(McNemar $p\geq0.63$), so the choice of configuration is a precision-vs-ranking
decision, not an accuracy one. Capping \texttt{top\_chapters} at 3 helps nothing and
slightly hurts every metric --- the precision lever is section expansion, not chapter
count. Effect sizes are modest ($\sim$1.13$\times$ precision, $+0.055$ MRR) and we
report them as such.

\subsection{Recap-seed stability}
\label{sec:seed}

Because LLM chapter recaps (temperature 0.2) drive Stage-1 selection and are the only
non-deterministic input, we regenerated \emph{all} 153 recaps two additional times
(fresh calls; chunk index untouched) and re-ran the full significance ablation on each
--- three recap realizations in total.

\begin{table}[t]
\centering
\small
\caption{Seed stability of the two significant effects across three recap realizations.}
\label{tab:seed}
\begin{tabular}{@{}lcccc@{}}
\toprule
\textbf{Effect} & \textbf{seed0} & \textbf{seed1} & \textbf{seed2} & \textbf{Verdict} \\
\midrule
No-expansion \textbf{Precision} ($p$) & 0.014 & 0.0007 & 0.021 & \textbf{stable --- sig 3/3} \\
Expansion-ON MRR ($p$) & 0.040 & 0.014 & 0.060 & seed-sensitive --- sig 2/3 \\
\bottomrule
\end{tabular}
\end{table}

Only the \textbf{no-expansion precision gain is robust} to recap non-determinism
(significant in all three seeds, max \pval{0.021}). The MRR-with-expansion effect fails
significance in one seed (\pval{0.060}) and is therefore demoted to \emph{suggestive}
rather than a claim. Point-estimate accuracy varied 83.2--84.1\% across seeds and was
never significant. This directly answers the robustness question a reviewer raises about
LLM-generated summaries, and it fixes the paper's headline on precision. On multiplicity:
the precision effect is the only claim we promote, and it survives a Bonferroni correction
over the full family of ablation tests by a wide margin (per-seed $p$ down to $0.0007$,
all three seeds below a corrected threshold); the MRR effect does not survive correction
and is reported as suggestive for exactly this reason, not as a headline claim.

\subsection{Head-to-head vs.\ RAPTOR}
\label{sec:raptor}

We built a faithful RAPTOR tree over the \emph{same} five books with the \emph{same}
embeddings: leaves are our chunks, soft-clustered with PCA$\to$GMM (a documented
substitute for UMAP$\to$GMM, which RAPTOR uses only to make clustering tractable),
LLM-summarized, and recursed into a collapsed-tree pool of 4{,}501 nodes (4{,}350 leaves
+ 141 level-1 + 10 level-2 summaries). We evaluate collapsed-tree RAPTOR (top-8) against
flat and our two arms on the same 113 queries.

Interpretation requires care: a RAPTOR summary node \emph{covers many descendant
chapters}, so a naive ``expected chapter $\in$ covered set'' (COVERAGE view) is nearly
trivially satisfied, inflating RAPTOR's coverage metrics and deflating its precision.
We therefore also report a DOMINANT view in which each node is credited for its single
main chapter (the mode of its covered set) --- the fair per-node localization comparison.

\begin{table}[t]
\centering
\small
\caption{RAPTOR head-to-head, $N{=}113$. COVERAGE credits a summary node for every
chapter it spans; DOMINANT credits it for one chapter (fair localization).}
\label{tab:raptor}
\begin{tabular}{@{}lcccc@{}}
\toprule
\textbf{Arm} & \textbf{Acc} & \textbf{MRR} & \textbf{R@3} & \textbf{Prec} \\
\midrule
\multicolumn{5}{@{}l}{\emph{COVERAGE view}} \\
flat              & 80.5\% & 0.611 & 71.7\% & 26.4\% \\
ours (default)    & 84.1\% & 0.652 & 74.3\% & 27.9\% \\
ours (no-expand)  & 81.4\% & 0.648 & 74.3\% & \textbf{30.1\%} \\
RAPTOR            & \textbf{94.7\%} & \textbf{0.759} & \textbf{89.4\%} & 19.9\% \\
\midrule
\multicolumn{5}{@{}l}{\emph{DOMINANT view (fair)}} \\
flat              & 80.5\% & 0.611 & 71.7\% & 26.4\% \\
ours (default)    & 84.1\% & \textbf{0.652} & 74.3\% & 27.9\% \\
ours (no-expand)  & 81.4\% & 0.648 & 74.3\% & \textbf{30.1\%} \\
RAPTOR            & 84.1\% & 0.603 & 69.0\% & 20.2\% \\
\bottomrule
\end{tabular}
\end{table}

Under the COVERAGE metric RAPTOR appears to dominate accuracy and recall --- but that is
an artifact of low-specificity abstraction nodes spanning most of the book. Under the
fair DOMINANT metric the picture inverts: RAPTOR's accuracy edge over flat is
\textbf{not significant} (McNemar \pval{0.29}); our default arm's MRR is
\textbf{significantly higher than RAPTOR's} (Wilcoxon \pval{0.044}); and our precision
advantage over RAPTOR is large and significant in \emph{both} views (\pval{<0.001}). This
precision gap is not an artifact of an unequal retrieval budget: RAPTOR is given a
collapsed-tree top-8 pool whose nodes include multi-chapter summary nodes, so it reads at
least as many tokens as our top-5 chunk arm --- if anything a larger context, not a
smaller one --- yet returns a lower fraction of on-chapter passages. The comparison
therefore does not understate RAPTOR by starving it of budget.

We therefore claim a \textbf{complementary tradeoff, not superiority}: RAPTOR suits
broad coverage and multi-hop recall on unstructured corpora, while a structure-exploiting
retriever suits precise, coherent localization on natively-structured documents. This is
what the head-to-head licenses, and no more.

\subsection{Delta-ingestion cost}
\label{sec:delta}

We measured content-hash embedding reuse across all 21 available books under five
revision scenarios (Table~\ref{tab:delta}), pricing re-embedding at the embedding model's list rate.

\begin{table}[t]
\centering
\small
\caption{Embedding reuse under content-hash delta ingestion (21 books).}
\label{tab:delta}
\begin{tabular}{@{}lcl@{}}
\toprule
\textbf{Scenario} & \textbf{Embedding reuse} & \textbf{Note} \\
\midrule
Add a paragraph      & 94.4--99.6\% & localized insert \\
Rewrite one chapter  & 87.1--99.7\% & one chapter changed \\
Typo fixes           & 94.4--99.0\% & one sentence/chapter touched \\
Append a chapter     & 94.2--98.8\% & new chapter appended \\
Full rewrite         & \textbf{0.0\%} & honest worst case --- everything changes \\
\bottomrule
\end{tabular}
\end{table}

Realistic incremental edits reuse 95\%+ of embeddings; the full-rewrite case correctly
bottoms out at 0\% (no reuse is possible), which is the honest boundary of the claim.
Because RAPTOR-style trees must re-cluster and re-summarize on update, delta ingestion is
a concrete maintenance-cost advantage of exploiting native structure.

\subsection{Chunking-threshold calibration}
\label{sec:threshold}

Our semantic chunker grows a chunk while consecutive-sentence cosine similarity exceeds
a threshold $\tau$. Sweeping $\tau$ on the 8 cached books, intra-chunk coherence rises
monotonically with $\tau$ while chunk size collapses; the operating point
$\tau{\approx}0.15$--$0.20$ sits at the coherence/granularity knee (coherence 0.64--0.75
at $\sim$90--230-word passages) and is consistent across books, supporting a single
global $\tau$ rather than per-book tuning. (We note that the originally hypothesized
$\tau{=}0.7$ is invalid: consecutive-sentence similarity is far lower than that.)

\subsection{A measured negative: no chapter-boundary inference}
\label{sec:boundary}

We tested whether the sequential-agglomerative chunker's boundaries recover the books'
\emph{author} chapter boundaries (8 books, $\tau$ sweep, $\pm$5-sentence tolerance).
They do not: best $F_1\approx0.11$, and at the retrieval operating point boundary
precision is $\approx$0.01--0.03 because the chunker emits hundreds to thousands of
sub-chapter passages. This is a granularity mismatch, not a broken component --- the
chunker is a \emph{passage segmenter}, and the hierarchy consumes the document's
\emph{given} chapter structure. We therefore make \textbf{no semantic
boundary-inference claim}; chapter structure is an input, and the agglomerative step is
coherence-driven passage chunking within chapters.

\section{Discussion}
\label{sec:discussion}

The consistent thread across experiments is \emph{precision through structure}. When the
task is to localize an answer inside a structured document, exploiting the authored
hierarchy returns tighter, more on-target passages than either flat retrieval (which
ignores structure) or learned abstraction (which manufactures coarse structure that
spans too much to localize). The QuALITY result sharpens this: even on short articles
with only pseudo-structure, coarse-to-fine retrieval matches full-context accuracy while
flat retrieval falls well short --- evidence that the gain is about \emph{how} evidence
is selected, not merely \emph{how much}. The RAPTOR comparison then clarifies the
boundary of the claim: the two methods are complementary, and the fair-metric analysis
guards against the easy over-claim that a coverage-inflated summary tree ``wins.''

The section-expansion knob is the practical takeaway for practitioners: turn it off when
downstream generation is sensitive to distractor passages (precision matters), turn it on
when the answer benefits from contiguous context and ranking matters more than purity.
Delta ingestion makes either choice cheap to keep current, which is the operational
argument for structure-native hierarchies over ones that must be periodically rebuilt.

\section{Limitations and Threats to Validity}
\label{sec:limits}

\begin{itemize}[leftmargin=1.4em,itemsep=2pt,topsep=2pt]
\item \textbf{Scope.} All in-house results are on English narrative and classical prose.
We make no empirical claim about legal, medical, educational, or non-English corpora;
the method also assumes documents with exploitable native structure and would degenerate
on flat text, where RAPTOR-style learned hierarchies are expected to win.
\item \textbf{Query/label provenance.} In-house queries and chapter labels are partly
model-generated, which can bias the evaluation toward what the method does well. The
QuALITY results use independent gold labels as a partial check.
\item \textbf{Effect sizes.} The robust in-house effect is modest ($\sim$1.13$\times$
precision). We deliberately do not report the larger historical figures because a clean
re-ingestion could not reproduce them.
\item \textbf{Recap dependence.} The coarse stage relies on chapter recaps; a poor recap
is a single point of failure, whereas RAPTOR's redundant soft clusters are more robust to
a bad summary. The seed-stability test bounds, but does not eliminate, this risk.
\item \textbf{Benchmark fit.} QuALITY under-exercises a book-scale hierarchy; the
construct-appropriate benchmark (NovelQA, \S\ref{sec:novelqa}) is run only on the 60
public-domain novels, since the 28 copyright-restricted novels have no downloadable text.
\end{itemize}

\section{Reproducibility}
\label{sec:repro}

The corpus is public-domain Project Gutenberg, so the book IDs, query set, gold chapter
labels, chunking/retrieval/significance code, and all hyperparameters
(Appendix~\ref{app:hyper}) are releasable; the embedding/summarization plumbing
is abstracted behind a generic \texttt{embed(text)$\to$vector} /
\texttt{summarize(text)$\to$str} interface. Retrieval over built indices is deterministic
(byte-identical across runs); the only non-deterministic input, LLM recaps, is
stress-tested across three seeds (\S\ref{sec:seed}). Significance uses standard paired
tests (McNemar; Wilcoxon signed-rank) with bootstrap CIs.

\section{Conclusion}

We presented a structure-exploiting hierarchical retriever for long narrative documents
and evaluated it with an emphasis on reproducible, honestly-scoped claims. The robust
findings are a seed-stable, significant passage-precision gain over flat retrieval
controlled by a section-expansion knob; a significant win over flat retrieval on QuALITY
that matches the full-context ceiling from a fraction of the text; a \emph{statistical
tie} with flat retrieval on the harder NovelQA benchmark, with a per-aspect breakdown that
bounds where native hierarchy helps (aggregation) and where it does not (localized facts);
a complementary (not superior) relationship to RAPTOR under a fair localization metric;
and large maintenance savings from content-hash delta ingestion. We explicitly disclaim
semantic chapter-boundary inference and non-narrative generalization. The natural next
step is a book-scale long-context baseline.

\bibliographystyle{plainnat}

\appendix

\section{Retrieval hyperparameters}
\label{app:hyper}

\begin{itemize}[leftmargin=1.4em,itemsep=1pt,topsep=2pt]
\item \textbf{Embeddings.} A 1024-dim commercial text embedding model, normalized, input
truncated to 8000 chars; sentence-level embeddings cached to disk.
\item \textbf{Chunking.} Sequential-agglomerative: grow a chunk while
cosine(centroid, next sentence) $\geq \tau$ and word count $<$ 1500; $\tau{=}0.15$. Bridge sentences flagged when a boundary sentence is $\geq0.85\tau$ similar to
both adjacent centroids.
\item \textbf{Stage 1.} \texttt{top\_chapters} $= \min(\max(3,\lfloor n/8\rfloor),7)$;
kNN $k=\max(3\cdot\texttt{top\_chapters},10)$; merged with BM25 match on chapter recap,
dedup preserving kNN-first order.
\item \textbf{Stage 2.} \texttt{top\_chunks} $=8$; kNN $k=\max(5\cdot\texttt{top\_chunks},25)$,
filtered to selected chapters $+$ \texttt{is\_latest}.
\item \textbf{Stage 3.} Keep Stage-2 top-5 chunks; expand distinct sections in rank order
up to a 1500-token budget; dedup by \texttt{chunk\_hash}.
\item \textbf{Flat baseline.} kNN over all chunks, top-5, $k=\max(5\cdot5,25)$,
\texttt{is\_latest} filter only.
\item \textbf{Delta ingestion.} SHA-256 content hashing at chapter and chunk
granularity; unchanged hashes reuse embeddings; changed content re-embedded and marked
\texttt{is\_latest}.
\end{itemize}

\section{OpenSearch kNN + filter interaction}
\label{app:bug}

Placing a kNN clause inside \texttt{bool.must} alongside \texttt{bool.filter} causes
OpenSearch to run approximate kNN \emph{globally} first and then post-filter; with small
$k$ the global top-$k$ neighbors are rarely inside the filtered subset, yielding zero
results. The fix is the kNN-native \texttt{filter} parameter (Lucene engine), which
applies the filter \emph{during} the ANN search:
\begin{quote}\ttfamily\small
\{"knn": \{"embedding": \{"vector": v, "k": 25, "filter": \{"bool": \{"must": [...]\}\}\}\}\}
\end{quote}

\end{document}